\documentclass[entropy,article,submit,oneauthor,pdftex,10pt,a4paper]{Definitions/mdpi} 
\usepackage{amsmath}
\usepackage{graphicx}
\usepackage{subfigure}
\usepackage{amsthm}
\usepackage{verbatim}
\usepackage{dcolumn}% Align table columns on decimal point
\usepackage{bm}% bold math
\usepackage{epsf}
\usepackage{color}
\usepackage{dsfont}
\usepackage{amssymb}
\usepackage{stmaryrd}
\newcommand\myeqref[1]{
	Eq. (\textup{\ref{#1}})
}
\newcommand{\bra}[1]{\left\langle #1\right|}
\newcommand{\ket}[1]{\left|#1\right\rangle}

\newcommand{\tr}[1]{\mathrm{tr}\left\{#1\right\}}
\newcommand{\ptr}[2]{\mathrm{tr_{#1}}\left\{#2\right\}}

\newcommand{\la}{\left\langle}
\newcommand{\ra}{\right\rangle}
\newcommand{\pd}{\partial}

\newcommand{\miU}[1]{\min_{U}{\left\{#1\right\}}}

\newcommand{\bla}{bla\\bla\\bla\\bla\\bla}

\newcommand{\mc}[1]{\mathcal{#1}}

\firstpage{1} 
\pubvolume{xx}
\issuenum{1}
\articlenumber{5}
\pubyear{2018}
\copyrightyear{2018}
\history{Received: date; Accepted: date; Published: date}

\Title{Quantum Euler relation for local measurements}

\Author{Akram Touil $^{1,*}$\orcidB{}, Kevin Weber $^{1,2}$\orcidA{} and Sebastian Deffner $^{1,3}$\orcidC{}}
\address{%
$^{1}$ \quad Department of Physics, University of Maryland, Baltimore County, Baltimore, MD 21250, USA\\
$^{2}$ \quad  Institute for Physical Science and Technology, University of Maryland, College Park, Maryland 20742, USA\\
$^{3}$ \quad Instituto de F\'isica `Gleb Wataghin', Universidade Estadual de Campinas, 13083-859, Campinas, S\~{a}o Paulo, Brazil\\
$^{*}$ \quad Correspondence: akramt1@umbc.edu
}

\corres{$^{*}$ Correspondence: akramt1@umbc.edu}

\abstract{In classical thermodynamics the Euler relation is an expression for the internal energy as a sum of the products of canonical pairs of extensive and intensive variables.  For quantum systems the situation is more intricate, since one has to account for the effects of the measurement back action. To this end, we derive a quantum analog of the Euler relation, which is governed by the information retrieved by local quantum measurements. The validity of the relation is demonstrated for the collective dissipation model, where we find that thermodynamic behavior is exhibited in the weak-coupling regime.}

\preto{\abstractkeywords}{\nolinenumbers}
\begin{document}

\section{Introduction}

Thermodynamics is a phenomenological theory studying the average behavior of heat and work~\cite{callen1998thermodynamics,huang2009introduction} originally developed to optimize heat engines~\cite{carnot1978reflexions}. Remarkably, this theory has been extended far beyond its origins and led to profound fundamental statements about nature, such as the second law linking the monotonicity of entropy to the arrow of time~\cite{huang2009introduction}, and the seminal role of information in physics~\cite{info1,info2}. Building on the success of this theory, \textit{quantum thermodynamics}~\cite{QT1,JPA_Goold} emerged as generalization of stochastic thermodynamics~\cite{stochastic1,stochastic2,stochastic3,QT1} to the quantum realm.

Quantum thermodynamics aims to ground our understanding of the universal laws and statements on a fundamental and genuine quantum level \cite{QT1}, which is set to have a direct impact on the development of new generation quantum technologies~\cite{PRXQuantum.1.020101,Dowling_Milburn}. Therefore, many recent studies focused on probing the role of quantum information as a thermodynamic resource~\cite{francica2020quantum,NJP_Korzekwa,ergo1,PLA_Giacomo,PRA_Funo,npjQI_Francica,Entropy_Mauro,PRL_Manzano,corr,PRE_Fusco,correlation1,correlation2,correlation3,touil2021second}, and deriving second law statements for quantum systems~\cite{touil2021second,seclaw,QT1}. However, to the best of our knowledge,  fundamental statements such as quantum generalizations of the Euler relation are still lacking.

In classical thermodynamics, the Euler relation is a fundamental statement relating the internal energy and the Clausius entropy,  with the volume and the particle number. This relation is derived using Euler's homogeneous function theorem~\cite{callen1998thermodynamics}, hence the name ``Euler relation''. The main goal of the present analysis is to derive a quantum analog of the Euler relation, that we dub \textit{the quantum Euler relation}. To this end, we have to relate the internal energy to thermodynamic quantities as well as the information output of quantum measurements. Consequently, our new relation connects the information gained from quantum measurements to inherent thermodynamic properties of a system. 

In the following, we focus on an established measure that quantifies the output information of a quantum measurement and its effects on the system we are probing. This measure was first introduced in the seventies~\cite{Groenewold1971}, when Groenewold proposed entropy reduction as a straightforward quantifier of information gain. Hereafter, we will simply refer to this quantifier as the \textit{information gain} ``$\mc{I}_g$''. Note that the information gain is generally different from the Holevo information or the Holevo bound ``$\chi$''~\cite{Holevo,nielsen2002quantum}, which is widely used in the literature  to quantify the classical information gained from measurements of bipartite quantum states. The discrepancy between these two information theoretic quantities will be made clear in our analysis.

\begin{figure}[h]
	\centering
	\includegraphics[width=0.715\textwidth]{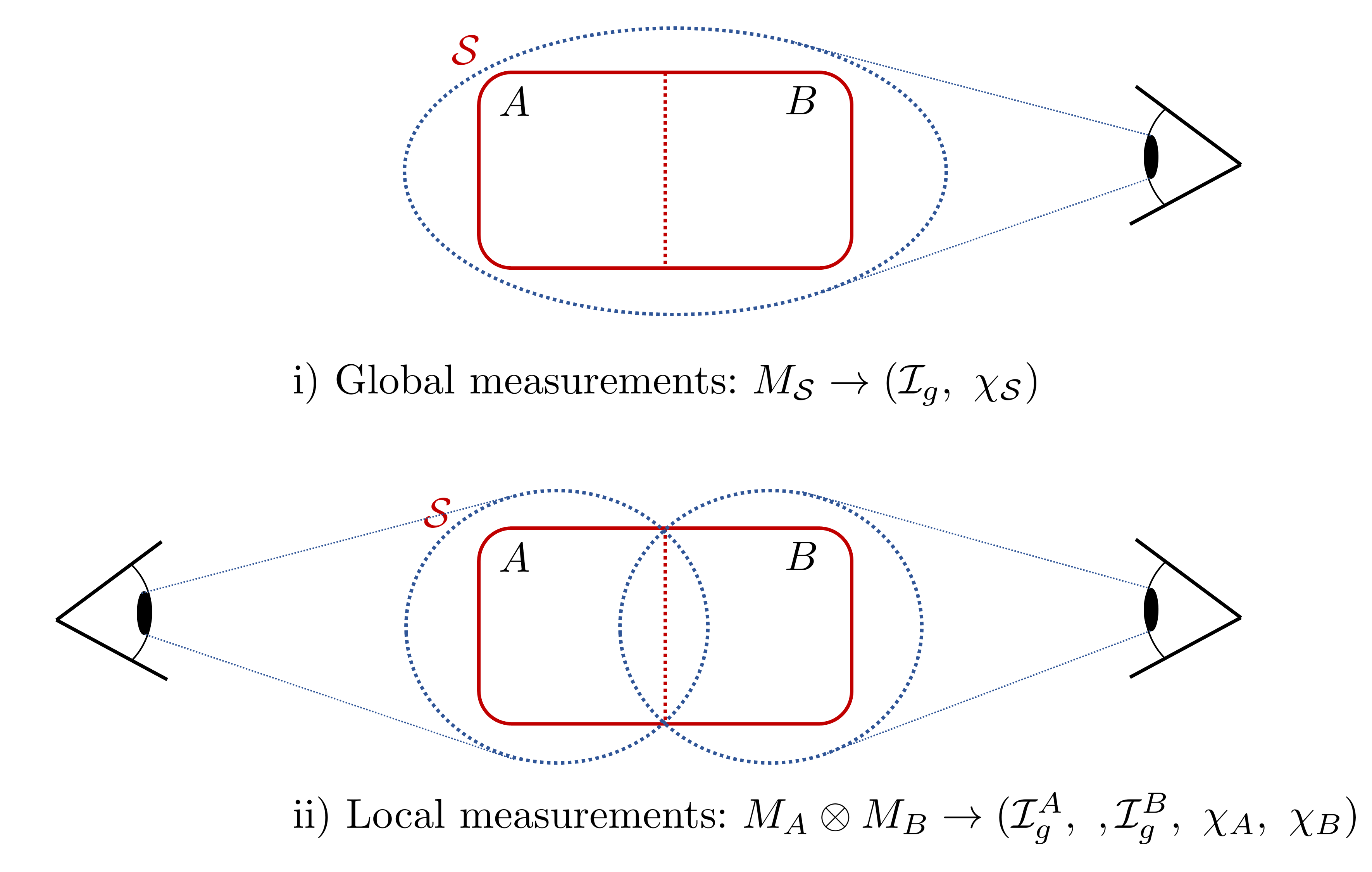}
	\caption{Sketch illustrating fundamentally different ways one can carry out quantum measurements: either (i) globally or (ii) locally. For global measurements, we are usually interested in quantities such as the information gain $\mc{I}_g$ and the Holevo bound $\chi_{\mc{S}}$. For local measurements, the direct quantities we measure are the local counterparts of $\mc{I}_g$ and $\chi_{\mc{S}}$: ($\mc{I}^{A}_g$, $\mc{I}^{B}_g$, $\chi_{A}$, $\chi_{B}$).}
	\label{illustration2}
\end{figure}

To obtain the quantum Euler relation, we consider the general setup of a quantum system $\mc{S}$ with two arbitrary partitions $A$ and $B$, in a Hilbert space $\mc{H}_{\mc{S}}=\mc{H}_{A} \otimes \mc{H}_{B}$. We start by exploring the connection between the information gain and bipartite correlations, classical and quantum, in the case of local measurements (cf. Fig.~\ref{illustration2}). In particular, we show that correlations between the two partitions of $\mc{S}$ directly hinder the amount of information gain accessible through measurements. Then, in Section~\ref{sec3}, we present our main results by first separating the information gain into classical and quantum contributions, and by deriving upper bounds on $\mc{I}_g$ as a function of the Holevo bound and thermodynamic quantities. 

We illustrate the tightness of the latter bounds in a collective dissipation model, where two qubits are collectively coupled to a thermal bath. Remarkably, by combining the aforementioned results and beyond our illustrative case study, we arrive at fundamental statements. The first, constitutes the quantum Euler relation involving the maximum work that can be extracted from the state of $\mc{S}$ (through unitary and cyclic operations) and the quantum contribution of the information gain $\mc{I}_g$. The second statement is a trade-off relation involving quantum correlations within the state of $\mc{S}$, and the entanglement that one of its partitions ($A$ or $B$) shares with the surrounding environment. 

\section{The information gain}

We start by introducing notions and notations before we will derive upper bounds on $\mc{I}_g$. This will clarify the effect of bipartite correlations (classical and quantum) on the amount of $\mc{I}_g$ that is accessible through measurements. 

\subsection{Positive operator-valued measures}

General quantum measurements~\cite{nielsen2002quantum} are described by a set of positive semi-definite operators, $\mc{M}=\left\{ M_n \right\}_{n \in \llbracket 1, \  d \rrbracket }$, where ``$n$'' labels the measurement outcome. These operators $M_n$ are called positive operator-valued measures or POVMs for short. They live on a $C^{*}$-algebra over the Hilbert space of the system being measured~\cite{nielsen2002quantum}, and satisfy the completness relation
\begin{equation}
\sum^{d}_{n=1} M^{\dagger}_n M_n= \mathbb{I}.
\end{equation}

For a general mixed state, characterized by a density matrix $\rho$, the $n$th measurement outcome occurs with a probability $p(n)$,
\begin{equation}
p(n)=\tr{M_n \rho M^{\dagger}_n},
\end{equation}
and the post-measurement state (after applying $M_n$) reads
\begin{equation}
\rho_n=\frac{M_n \rho M^{\dagger}_n}{p(n)}.
\end{equation}
Additionally, by considering the set of measurements $\mc{M}$, the end state of the measurement process reads $\mc{M}(\rho)=\sum^{d}_{n=1} p(n) \rho_n$,  which simplifies to $\mc{M}(\rho)=\sum^{d}_{n=1} M_n \rho M^{\dagger}_n$.  Here, we are implicitly considering efficient measurements where each measurement $M_n$ is represented by a single Kraus operator~\cite{efficient1,efficient2,efficient3,lindblad1972entropy,lindblad1973entropy,Groenewold1971,infogain}, as opposed to the case of weak~\cite{Qdemon} or inefficient measurements $M_n(\rho)=\sum_i M_{n,i} \rho M^{\dagger}_{n,i}$. Previous studies~\cite{efficient1,efficient2,efficient3,eff4,lindblad1972entropy,lindblad1973entropy} have shown that the information gain $\mc{I}_g$ is positive for all efficient measurements.

\subsection{Maximal information gain}

To probe the effect of correlations on the accessible information gain, we now derive two general upper bounds on $\mc{I}_g$. We begin by considering arbitrary POVMs, applied on a bipartite quantum system $\mc{S}$ living on the Hilbert space $\mc{H}_{\mc{S}}=\mc{H}_{A}\otimes \mc{H}_{B}$, with partitions $A$ and $B$, and dimension $d_{\mc{S}}=d_A d_B$. The information gain $\mc{I}_g$ can be written as~\cite{Groenewold1971}
\begin{equation}
\mc{I}_g = S(\rho_{AB})-\sum_{n}p_nS(\rho_n),
\label{def1}
\end{equation}
where $S(\rho)=-\tr{\rho \ln(\rho)}$ is the von Neumann entropy, and the density matrices $\rho_{AB}$ and $\rho_n$ represent the pre-measurement and post-measurement (with probability $p_n$) states, respectively.

From the definition of the quantum mutual information between the two partitions $A$ and $B$,
\begin{equation}
\mc{I}(A:B)= S(\rho_A)+S(\rho_B)-S(\rho_{AB}),
\label{MI}
\end{equation}
and for any choice of POVMs, we arrive at a trivial bound by recognizing that the second term in the expression of the information gain (cf.~\myeqref{def1}) is negative, hence
\begin{equation}
\begin{split}
\mc{I}_g \leq \ln(d_{\mc{S}})- \mc{I}(A:B).
\label{trivial}
\end{split}
\end{equation}
The bound in \myeqref{trivial} shows that the information contained in the correlations between $A$ and $B$ directly hinders the amount of accessible information gain. In other words, the higher the value of the mutual information $\mc{I}(A:B)$ the less information gain we can access through measurements.

Now, considering general, local POVMs~\footnote{These type of measurements describe more realistic settings~\cite{nielsen2002quantum,infogain,Qdemon,localm}, where one usually considers a quantum system interacting with an environment/ancilla. In such scenarios, measurements of a physical observable are either applied on the system (cf.~\cite{Qdemon}, $M_n= M^{A}_{n} \otimes \mathbb{I}_B$) or simultaneously on system and environment (cf.~\cite{localm}, $M_n=M_{n_A} \otimes M_{n_B} $). More specifically,  to evaluate the work applied on the system or the heat exchanged with the environment, these measurements take the form of local projective energy measurements. The latter will be the main focus of the following section where we explore the thermodynamic significance of the information gain.} applied on $\rho_{AB}$, of the form $M_n=M_{n_A} \otimes M_{n_B}$ (cf. Fig.~\ref{illustration2}), we get
\begin{equation}
\begin{split}
\mc{I}_g &= S(\rho_{A})+S(\rho_{B})-\mc{I}(A:B) -\sum_np_n \left(S(\rho^{n}_{A})+S(\rho^{n}_{B})-\mc{I}^{n}(A:B)\right),\\
&= \left(S(\rho_{A})-\sum_np_n S(\rho^{n}_{A})\right)+\left(S(\rho_{B})-\sum_np_n S(\rho^{n}_{B})\right) - \mc{I}(A:B) + \sum_np_n \mc{I}^{n}(A:B),\\
\end{split}
\end{equation}
hence,
\begin{equation}
\mc{I}_g = \mc{I}^{A}_g+\mc{I}^{B}_g - \mc{I}(A:B) + \sum_np_n \mc{I}^{n}(A:B),
\label{equ1}
\end{equation}
where $\mc{I}^{A}_g$ and $\mc{I}^{B}_g$ refer to the local information gain on each partition.  We have~\cite{generalm1,generalm2}
\begin{equation}
\sum_np_n \mc{I}^{n}(A:B) \leq \mc{I}(A:B),
\label{ineqq1}
\end{equation}
which directly implies the inequality
\begin{equation}
\mc{I}_g \leq \mc{I}^{A}_g+\mc{I}^{B}_g,
\label{result}
\end{equation}
that is reminiscent of the subadditivity of the von Neumann entropy~\cite{nielsen2002quantum}.  In fact, from the above inequality we infer that, due to correlations between $A$ and $B$, the total information gain we can access locally (from each partition) is greater or equal to the information gain of the composite system, and the more correlations between the partitions the less information gain we can access. 

In addition it is interesting to note that from the expression in~\myeqref{equ1} we can define a new quantity $\mc{I}_l=\mc{I}(A:B) - \sum_np_n \mc{I}^{n}(A:B)$, which represents the amount of total correlations lost as a consequence of measuring $\mc{S}$. Note that, following~\myeqref{ineqq1}, the quantity $\mc{I}_l$ is positive for local measurements. This quantity might prove useful in quantifying the invasiveness of quantum measurements, i.e., their effect on the correlations in a bipartite state. Finally, note that since the quantum mutual information does not increase under general local measurements we consider that the post measurement states have support on orthogonal subspaces~\cite{nielsen2002quantum}. In particular we  forgo classical communication between the two partitions, and hence through general local measurements alone the quantum mutual information does not increase~\cite{entanglement}.

\paragraph{Illustrative example:} Consider an arbitrary bipartite state described by a density matrix $\rho_{AB}$, and von Neumann measurements~\cite{nielsen2002quantum} $M_n=\Pi_{n_A} \otimes \Pi_{n_B}$ such that $\Pi_{n_A}$ ($\Pi_{n_B}$) are rank-one projectors acting on subsystem $A$ ($B$). In this scenario,~\myeqref{result} becomes
\begin{equation}
S(\rho_{AB}) \leq S(\rho_{A})+S(\rho_{B}).
\end{equation}
This is nothing but the subadditivity of the von Neumann entropy~\cite{nielsen2002quantum}, and it reflects the fact that the quantum mutual information is positive $\mc{I}(A:B) \geq 0$. Additionally, the inequality is saturated if and only if the partitions of the composite system are uncorrelated, i.e. $\rho_{AB}=\rho_{A} \otimes \rho_{B}$.

It is worth emphasizing that the subadditivity of the information gain~\myeqref{result}, is valid for general local measurements such that the post-measurement states have support on orthogonal subspaces~\cite{nielsen2002quantum}. In scenarios where these local measurements are represented by rank-one projectors we recover the subadditivity of the von Neumann entropy.

\subsection{The Holevo bound}

However, before we continue with local measurements on a single partition and derive our main results, it is crucial to consider the subtle difference between the information gain and the Holevo bound, first. The Groenewold information gain is directly related to the Holevo bound $\chi_{\mc{S}}$ (defined through measurement $\mc{M}$ on the composite system $\mc{S}$) via~\cite{infogain},
\begin{equation}
\mc{I}_g + \Delta = \chi_{\mc{S}},
\end{equation}
where $\Delta= S(\mc{M}(\rho_{AB}))-S(\rho_{AB})$ represents the change in entropy caused by the measurement. This quantity is also referred to as the entropy cost of a measurement~\cite{infogain}, and, as the name suggests, it quantifies the change in entropy between the pre- and post- measurement states. Depending on the type of measurements we carry out, the quantity $\Delta$ can be either positive or negative. For instance, it is straightforward to infer that for all projective measurements $\Delta \geq 0$~\cite{nielsen2002quantum}. 

\paragraph{Illustrative example:} To illustrate the discrepancy between these quantities we consider the same example we studied above. Namely, for von Neumann measurements $M_n=\Pi_{n_A} \otimes \Pi_{n_B}$ composed of rank-one projectors on each subsystem. Now, we further assume that the arbitrary state $\rho_{AB}$ has coherences in the basis determined by the latter projectors. Thus, the Holevo bound becomes
\begin{equation}
	\chi_{\mc{S}} = S(\mc{M}(\rho_{AB}))-\sum_{n}p_nS(\rho_n),
\end{equation}
which implies
\begin{equation}
	\chi_{\mc{S}}= S(\mc{M}(\rho_{AB}))-0=S(\rho^{\text{diag}}_{AB}),
\end{equation}
where $\rho^{\text{diag}}_{AB}$ is the incoherent state of $\mc{S}$, generated by removing the off-diagonal elements of $\rho_{AB}$. On the other hand, the information gain is equal to the von Neumann entropy of $\mc{S}$, $\mc{I}_g = S(\rho_{AB})$. Therefore, it is straightforward to get the strict inequality 
\begin{equation}
	\mc{I}_g < \chi_{\mc{S}},
\end{equation}
since the difference between the two quantities ($\chi_{\mc{S}}-\mc{I}_g$) is exactly the relative entropy of coherence~\cite{relative} in the basis determined by the rank-one projectors.

\section{Projective local measurements}
\label{sec3}

We are now positioned to relate the information gain to thermodynamic quantities, and derive  fundamental statements in quantum information theory as well as in quantum thermodynamics. To this end, we focus on the case of local projective energy measurements~\cite{PhysRevA.86.044302,localm} such that
\begin{equation}
	\mc{M}=\left\{\mathbb{I}_A \otimes |E^{n}_B \rangle \langle E^{n}_B|\right\}_{n \in \llbracket 1, \  d_B \rrbracket }=\left\{ M_n \right\}_{n \in \llbracket 1, \  d_B \rrbracket },
\label{bmeas}
\end{equation}
where $|E^{n}_B \rangle$ is the $n$th energy eigenstate of the self Hamiltonian of subsystem $B$, and $d_B$ is the dimension of its Hilbert space. Moreover, we assume the state $\rho_{AB}$ to be locally thermal (on both partitions) with inverse temperature $\beta$, i.e. $\rho_A=\exp(-\beta H_A)/Z_A$ and $\rho_B=\exp(-\beta H_B)/Z_B$, while $A$ and $B$ share correlations such that the global state  $\rho_{AB}$ is not necessarily thermal. Similar scenarios were discussed in the literature~\cite{touil2021second,corr} to assign thermodynamic value to quantum correlations. More specifically, we have shown in Ref.~\cite{touil2021second}  that the maximum work that can be extracted from a bipartite quantum state ``$\rho_{AB}$'', under unitary and cyclic operations, is directly related to the quantum mutual information ``$\mc{I}(A:B)$'', which will be crucial later on in our analysis.

It is a simple exercise to show that we have
\begin{equation}
\left(\forall n \in \llbracket 1, \  d_B \rrbracket \right); \ \ S(\rho^{n}_B)=\mc{I}^{n}(A:B)=0. 
\end{equation}
Therefore, \myeqref{equ1} simplifies to
\begin{equation}
\begin{split}
\mc{I}_g &= \mc{I}^{A}_g+S(\rho_B) - \mc{I}(A:B),\\
 &= \chi_B+S(\rho_B) - \mc{I}(A:B),
\end{split}
\label{result1}
\end{equation}
where $\chi_B$ is the Holevo information obtained by projective measurements on $B$~\footnote{In this case, we have $\mc{I}^{A}_g=\chi_B$ since we apply local measurements on $B$. The derivation directly follows from the cyclic property of the partial trace for the local measurements we are considering (cf.~\myeqref{bmeas}). More specifically, we have
	\begin{equation}
			\chi_{B}=S(\ptr{B}{\mc{M}(\rho_{AB})})-\sum_np_n S(\rho^{n}_{A})= S(\rho_{A})-\sum_np_n S(\rho^{n}_{A}),
\end{equation}
which implies  $\mc{I}^{A}_g=\chi_B$.} (cf.~\myeqref{bmeas}), and it reflects the information we gain about the partition $A$ by measuring subsystem $B$. From the above equality we can separate the information gain into classical and quantum contributions, and we can relate $\mc{I}_g$ to the maximum work extractable from $\rho_{AB}$. As a consequence of combining both results we will be able to derive fundamental statements for $\mc{I}_g$.

\subsection{Quantum and classical contributions}

Now exploiting the Koashi-Winter relation~\cite{koashi}, \myeqref{result1} can be written as
\begin{equation}
\mc{I}_g= \chi_B+\chi^{\text{max}}_A+E(B:C) - \mc{I}(A:B),
\label{p1}
\end{equation}
for a subsystem ``$C$'' that purifies the state of $\mc{S}= A \cup B$. This subsystem can be regarded as the environment for the system of interest $\mc{S}$, such that the state of the quantum universe ``$ABC$'' is pure. We emphasize that we are implicitly adopting the realistic setting of a system of interest $\mc{S}$ open to an environment ``$C$'', with which it can share quantum as well as classical correlations. Moreover, $\chi^{\text{max}}_A$ refers to the Holevo bound under optimal measurements on $A$, and ``$E(B:C)$'' is the entanglement of formation~\cite{eof} which represents a generalization of entanglement entropy to mixed states. To elaborate, the entanglement of formation ``$E(B:C)$'' quantifies the entanglement in the mixed state $\rho_{BC}$, since the entanglement entropy is no longer a viable measure of entanglement for such states~\cite{nielsen2002quantum}.

The expression in~\myeqref{p1} is further simplified by noting that the quantum mutual information can be decomposed into classical and quantum contributions~\cite{dis1,LH01}. In fact, the quantum contribution is known as quantum discord which is defined as the difference between the mutual information and the Holevo bound~\cite{dis1}
\begin{equation}
\mc{D}_{\mc{A}}=\mc{I}(A:B)-\chi^{\text{max}}_A,
\end{equation}
hence it quantifies purely the quantum correlations in a bipartite state. From this definition, we get the equality
\begin{equation}
	\mc{I}_g= \chi_B+(E(B:C)-\mc{D}_A),
	\label{mainb}
\end{equation}
which expresses the information gain, on the composite state of $\mc{S}$, in terms of classical and quantum contributions. The classical contribution is the Holevo information $\chi_B$, and the quantum contribution is represented by the difference between the entanglement of formation $E(B:C)$ and quantum discord $\mc{D}_A$, which we will refer to as the quantum information gain ``$\mc{I}^{Q}_g$''. This is a crucial result that will lead to interesting conclusions, as we will explore in the remainder of our analysis. Namely, once we look at the problem from a thermodynamics perspective through the notion of \textit{ergotropy}~\cite{allahverdyan2004maximal} (aka the maximum extractable work under unitary and cyclic operations).

\subsection{Maximum extractable work}

The maximum amount of work that can be extracted from the composite state $\rho_{AB}$, through unitary and cyclic operations, is referred to as the ergotropy ``$\mc{E}$''~\cite{allahverdyan2004maximal,touil2021second,sone2021quantum}. Consider a quantum system with Hamiltonian $H=\sum_{i=1}^{d}\varepsilon_{i}\left|\varepsilon_{i}\right\rangle\left\langle\varepsilon_{i}\right|$  and quantum state $\rho=\sum_{j=1}^{d} r_{j}\left|r_{j}\right\rangle\left\langle r_{j}\right|$, such that $\varepsilon_{i} \le \varepsilon_{i+1}$ and $r_{j} \geq r_{j+1}$. The ergotropy is calculated by performing an optimization over all possible unitary operations to achieve a final state that has the minimum average energy with respect to $H$,
\begin{equation}
	\mathcal{E}(\rho)=\tr{H\rho}-\miU{\tr{H U\rho U^{\dag}}}=\tr{H(\rho-P_{\rho})},
	\label{gergo}
\end{equation}
where $P_{\rho}\equiv\sum_{k} r_{k}\ket{\varepsilon_{k}}\bra{\varepsilon_{k}}$ is called the \emph{passive state}. An equivalent expression reads
\begin{equation}
	\mathcal{E}(\rho)=\sum_{i, j} r_{j}\varepsilon_{i}\left(|\la r_j | \varepsilon_i \ra|^2-\delta_{ij}\right).
\end{equation}
Assuming the passive state $P_{\rho}$ is not thermal, we can still extract work out of it, if we have access to, and ability to globally act upon, multiple copies of this state. This is captured by the bound ergotropy $\mc{E}_b$~\cite{niedenzu2019concepts,touil2021second},
\begin{equation}
	\label{eq:bound}
	\mc{E}_b(\rho)= \tr{(P_{\rho}-P^{\text{th}}_{\rho}) H},
\end{equation}
where $P^{\text{th}}_{\rho}$ is a thermal state such that $S(P_{\rho})=S(P^{\text{th}}_{\rho})$. In other words, $P^{\text{th}}_{\rho}$ is the thermal state associated with the passive state $P_{\rho}$.

From the  definition of the ergotropy and using the result of Ref.~\cite{touil2021second}, namely the relationship between the quantum mutual information and the ergotropy, $\beta \mc{E} \leq \mc{I}(A:B)$, and \myeqref{result1}, we obtain
\begin{equation}
	\mc{I}_g \leq \chi_B+ \beta \left(\la H_B \ra - \mc{E}-F_B \right).
	\label{ineq1}
\end{equation}
In the above inequality, $F_B=- \left(1/\beta \right) \log(Z_B)$ is the Helmholtz free energy. Furthermore, considering the bound ergotropy $\mc{E}_b$, where $\mc{E}_G=\mc{E}+\mc{E}_b$ is the global ergotropy, and using the general inequality $	\beta \mc{E}_G \leq \mc{I}(A:B)$ from Ref.~\cite{touil2021second}, we arrive at the tighter bound
\begin{equation}
	\mc{I}_g \leq \chi_B+ \beta \left(\la H_B \ra -\mc{E}_G-F_B \right).
\label{ineq2}
\end{equation}

In fact, the above inequality is saturated for a large class of states. This will be exemplified in the following section through a collective dissipation model. For the class of states where the above inequality is saturated, and using \myeqref{mainb}, we get
\begin{equation}
E(B:C)-\mc{D}_A = \beta \left(\la H_B \ra -\mc{E}_G-F_B \right).
\label{mainb2}
\end{equation}
This result can be interpreted as follows. First, it represents a trade-off relation between the quantum correlations present in $\mc{S}$ (quantified by $\mc{D}_A$) and the entanglement between a partition of $\mc{S}$ ($B$ in this case) and the surrounding environment ($C$). This is complementary to the Koashi-Winter relation, which represents a trade-off between classical information and entanglement. Second, simply by rearranging the terms, we obtain a quantum Euler relation involving ergotropy, the free energy, and the quantum correlations within $\mc{S}$ as well as the entanglement that $\mc{S}$ shares with its surroundings, namely
\begin{equation}
\la H_B \ra = \mc{E}_G+F_B+\mc{I}^{Q}_g/\beta.
\label{mainb3}
\end{equation}
Equation~\eqref{mainb3} constitutes our main result. It relates the quantum information gained from local measurements to inherent thermodynamic properties of a bipartite system.

\section{Collective dissipation}

After having established the conceptual framework and the main results, this part of the analysis is dedicated to an illustrative case study. In particular, we illustrate the tightness of the bounds derived in~\myeqref{ineq1} and~\myeqref{ineq2}. To this end, we consider two-qubit X-shape density matrices of the form
\begin{equation}\label{xshape}
	\rho(t)=
	\begin{pmatrix}
		\rho_{11} & 0 & 0 & \rho_{14} \\
		0 & \rho_{22} & \rho_{23} & 0 \\
		0 & \rho_{23}^* & \rho_{33} & 0 \\
		\rho_{14}^* & 0 & 0 & \rho_{44} \\
	\end{pmatrix}\,.
\end{equation}
For such matrices, taking a partial trace on either qubits~\footnote{Here, the composite system $\mc{S}$ is described by the density matrix $\rho(t)$. Therefore, the partitions $A$ and $B$ are single qubits.} results in a thermal state with inverse temperature $\beta$. These X-shape density matrices are ubiquitous in the literature~\cite{x_states}, as they can be found in Pauli channels~\cite{xshape}, collective dephasing models for two-qubit systems~\cite{NJP_Carnio}, or in 1-D spin chains that exhibit $\mathbb{Z}_2$ symmetry~\cite{RMP_Amico,Sarandy,Entropy_Baris}.

Such X-states are characteristic for the \textit{collective dissipation model} where a two-qubit system is collectively coupled to a thermal bath with inverse temperature $\beta_e$. The master equation of the dynamics reads~\cite{PRA_Angsar,PRA_Latune,Gross1982,Stephen,Lehmberg,JPhysB_Damanet}
\begin{equation}
	\frac{\pd \rho}{\pd t} = -\frac{i}{\hbar}[(H_0+H_d), \rho]+D_-(\rho)+D_+(\rho),
\end{equation}
where $H_0\!=\!\hbar\omega(\sigma_1^+\sigma_1^-+\sigma_2^+\sigma_2^-)$ and $H_d\!=\!\hbar f(\sigma_1^+\sigma_2^-+\sigma_2^+\sigma_1^-)$ are the self-Hamiltonian of the whole system and the interaction Hamiltonian between the qubits, respectively. The $\sigma^{+}_i$ and $\sigma^{-}_i$ are the usual Pauli raising and lowering operators, respectively. Additionally,
\begin{equation}
	\begin{split}
		D_-(\rho) &=\sum\limits_{i,j=1}^2\gamma_{ij}\,(\bar{n}+1)(\sigma_j^-\rho\sigma_i^+-\frac{1}{2}\{\sigma_i^+\sigma_j^-, \rho\}), \\ 
		D_+(\rho) &=\sum\limits_{i,j=1}^2\gamma_{ij}\,\bar{n}(\sigma_j^+\rho\sigma_i^--\frac{1}{2}\{\sigma_i^-\sigma_j^+, \rho\}).
	\end{split}
\end{equation}
Here, $\bar{n}\!=\![\exp(\beta_e\omega)-1]^{-1}$ is the mean number of photons at the temperature of the environment $\beta_e$, and $\gamma_{ij}$ are the spontaneous decay rates. One can show that there is no unique fixed point for this model and the steady state $\rho_{ss}$ depends on the initial state $\rho_0$. In fact, the analytic expression for the X-shape steady states~\cite{touil2021second} is
\begin{equation}
\begin{split}
	\rho_{ss}&= (1-c)\ket{\psi_-}\bra{\psi_-} \\
	& +cZ_+^{-1}\left(\beta_e\right)\left(\exp(-2\omega\beta_e)\ket{\psi_{ee}}\bra{\psi_{ee}}+\exp(-\omega\beta_e)\ket{\psi_+}\bra{\psi_+}+\ket{\psi_{gg}}\bra{\psi_{gg}}\right),
\end{split}
\label{steady}
\end{equation}
where, $\ket{\psi_{gg}}\!=\!\ket{gg}$, $\ket{\psi_{ee}}\!=\!\ket{ee}$, $\ket{\psi_{\pm}}\!=\!\ket{ge}\pm\ket{eg}/\sqrt{2}$, $c\!=\!\bra{\psi_{gg}}\rho_0\ket{\psi_{gg}}+\bra{\psi_{ee}}\rho_0\ket{\psi_{ee}}+\bra{\psi_+}\rho_0\ket{\psi_+}$, and $Z_+\left(\beta_e\right)\!=\!1+\exp(-\omega\beta_e)+\exp(-2\omega\beta_e)$. It is instructive to realize that the parameter $c$ plays the role of an effective coupling constant. For $c=1$ the stationary state is thermal and the usual thermodynamic behavior is recovered. In particular,  in this case the stationary state is fully independent of the initial preparation, or in other words, the stationary state carries no memory of the initial state. In stark contrast, for $c=0$ the stationary state is strongly dependent on the initial preparation. Thus, one would expect thermodynamic statements to be only valid for $c\simeq 1$, whereas only weaker statements can hold for arbitrary $c$.

Also note that the inverse temperature $\beta$ of each partition/qubit is directly related to the bath temperature and we have
\begin{equation}
	\beta=\frac{1}{\omega}\ln\left[\frac{1+2\cosh(\beta_e\omega)+2c\sinh(\beta_e\omega)}{1+2\cosh(\beta_e\omega)-2c\sinh(\beta_e\omega)}\right].
\end{equation}
For more technical details on the collective dissipation model, interested readers can refer to Refs.~\cite{ergo1,touil2021second}. 

Here we continue by illustrating \myeqref{ineq1}. To this end, we  plot the right and left hand sides of the inequality for $\beta_e=10$ and $\omega=1$. In this limit, the ergotropy can be expressed analytically~\cite{ergo1}
\begin{equation}
\begin{split}
(\forall c \in [0,\frac{1}{2}]); \ \ \mc{E}&=1-2c,\\
(\forall c \in [\frac{1}{2},1]); \ \ \mc{E}&=0.
\end{split}
\label{ergotropyf1}
\end{equation}
\begin{figure}[h!]
	\centering
	\includegraphics[width=0.78\textwidth]{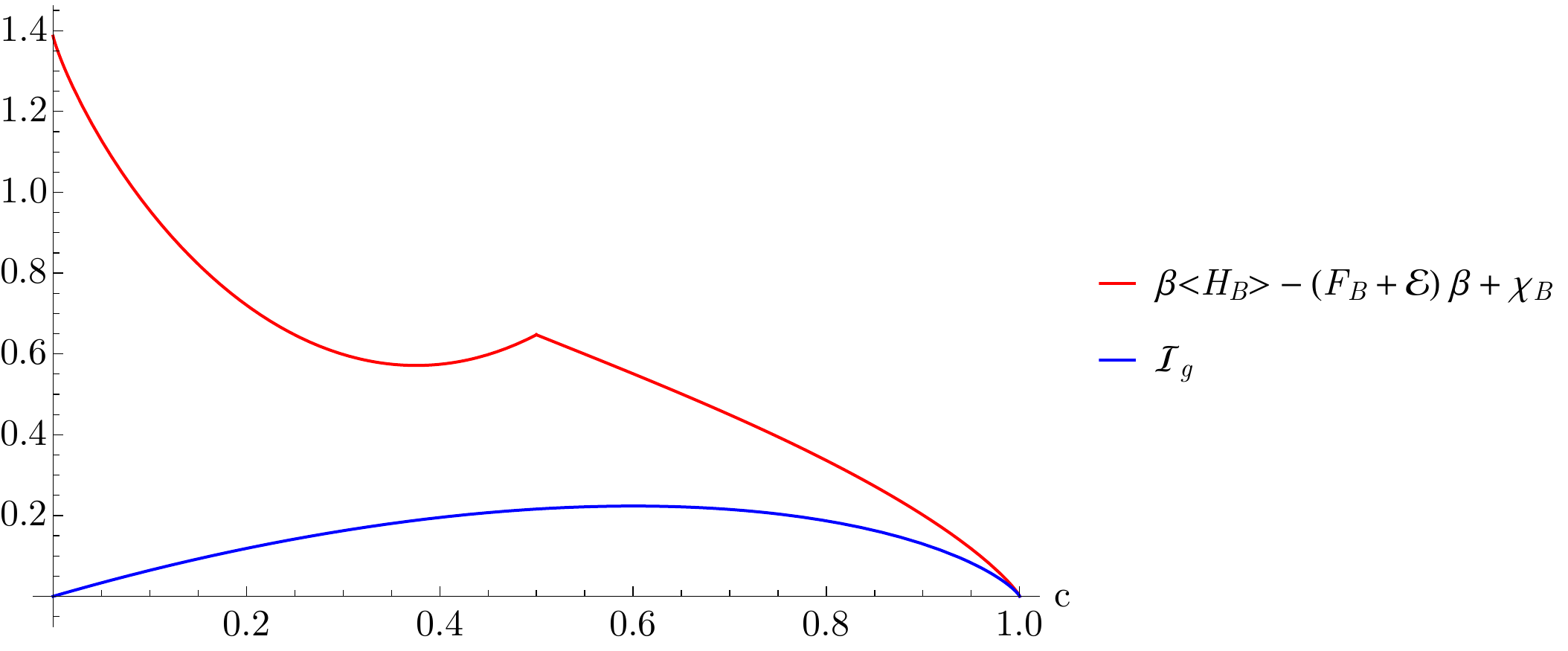}
	\caption[]{\label{bound1}Plots of the information gain $\mc{I}_g$ and the right hand side of \myeqref{ineq1} as a function of $c$.}
\end{figure}
\begin{figure}[h!]
	\centering
	\includegraphics[width=0.78\textwidth]{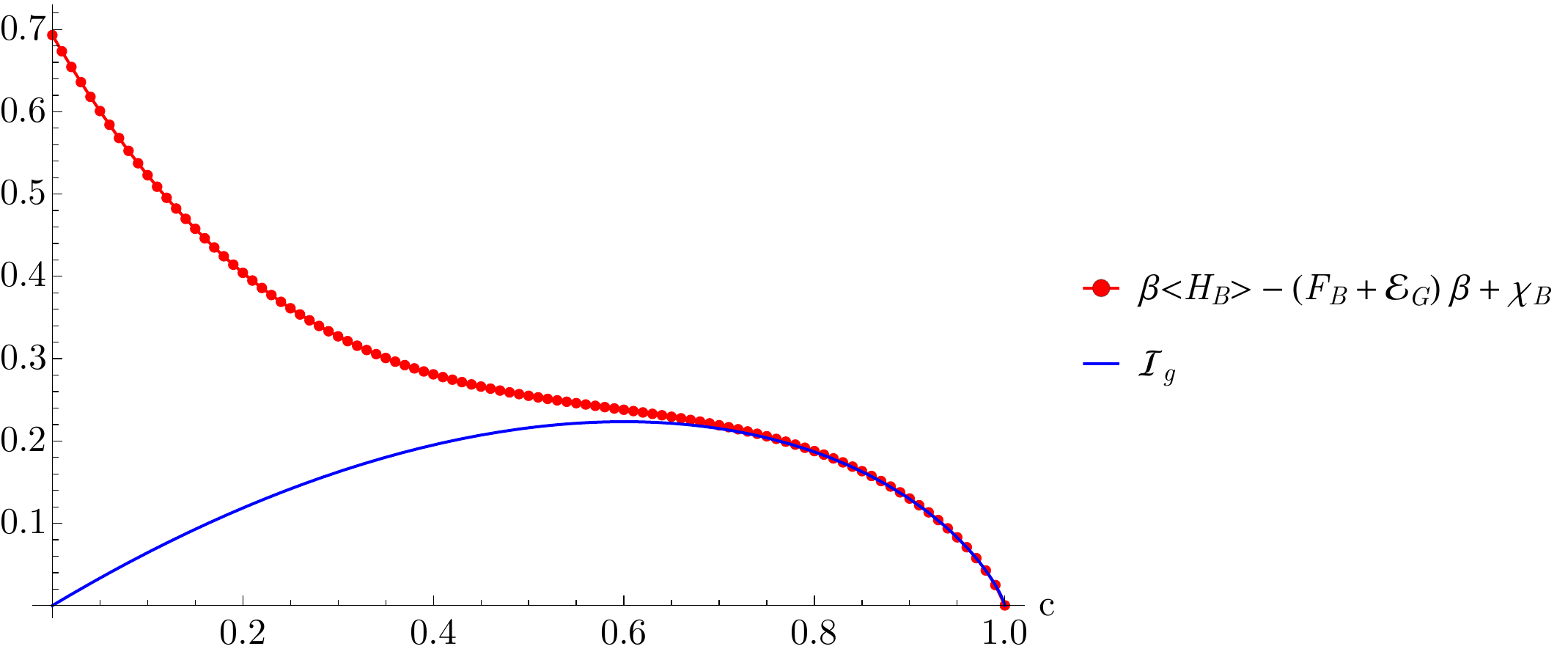}
	\caption[]{\label{bound2}Plots of the information gain $\mc{I}_g$ and the right hand side of \myeqref{ineq2} as a function of $c$.}
\end{figure}

From Figure.~\ref{bound1}, we observe that the upper bound is not a tight bound for almost all values of $c$, this is due to the fact that the mutual information is not a sharp upper bound to the ergotropy as observed in Ref.~\cite{touil2021second}. Additionally, the kink in the red curve at the value $c=0.5$ is a direct consequence of the functional form of the ergotropy, cf.~\myeqref{ergotropyf1}. For a tighter bound we illustrate~\myeqref{ineq2} in Figure.~\ref{bound2}, and we observe that the upper bound is saturated for all steady states where roughly $c \in [0.7,1]$.

The collective dissipation model we use in this section is a perfect example of an experimentally relevant scenario~\cite{RMP_Amico,Sarandy,Entropy_Baris} where the inequality presented in \myeqref{ineq2} is saturated. Consequently, this model shows the existence of a class of states where the statements in \myeqref{mainb2} and \myeqref{mainb3} are valid, which can be qualitatively explained as follows. Recall that $c$ quantifies the information the steady state carries about the initial state of the dynamics (cf.~\myeqref{steady}). Therefore, as $c$ gets closer to one we recover the limit where thermodynamic expressions such as the Euler relation (i.e. Equation~(\ref{mainb3})) are valid. Beyond this limit, we have the general inequality
\begin{equation}
	\la H_B \ra \geq \mc{E}_G+F_B+\mc{I}^{Q}_g/\beta.
\end{equation}

\section{Concluding remarks}

In the present analysis, we studied the information gain $\mc{I}_g$, originally proposed by Groenewold, to derive the quantum analog of the Euler relation. More specifically, we studied the relationship between $\mc{I}_g$ and bipartite total correlations (quantum and classical) which led to the quantum Euler relation as well as a general trade-off statement. First, we started by showing that total correlations hinder the amount of information gain $\mc{I}_g$ that can be accessed through measurements. We then presented the direct connection between the Holevo bound and $\mc{I}_g$. In Section~\ref{sec3}, we showcased our main results, where we considered bipartite states that are locally thermal. In particular, we proved that the information gain can be separated into quantum and classical contributions. Additionally, we derived upper bounds on $\mc{I}_g$ as a function of the Holevo bound, the average and free energies, as well as the ergotropy. The aforementioned inequalities were illustrated in the experimentally relevant example of a two-qubit system collectively coupled to a thermal bath.

\begin{figure}[h]
	\centering
	\includegraphics[width=0.92\textwidth]{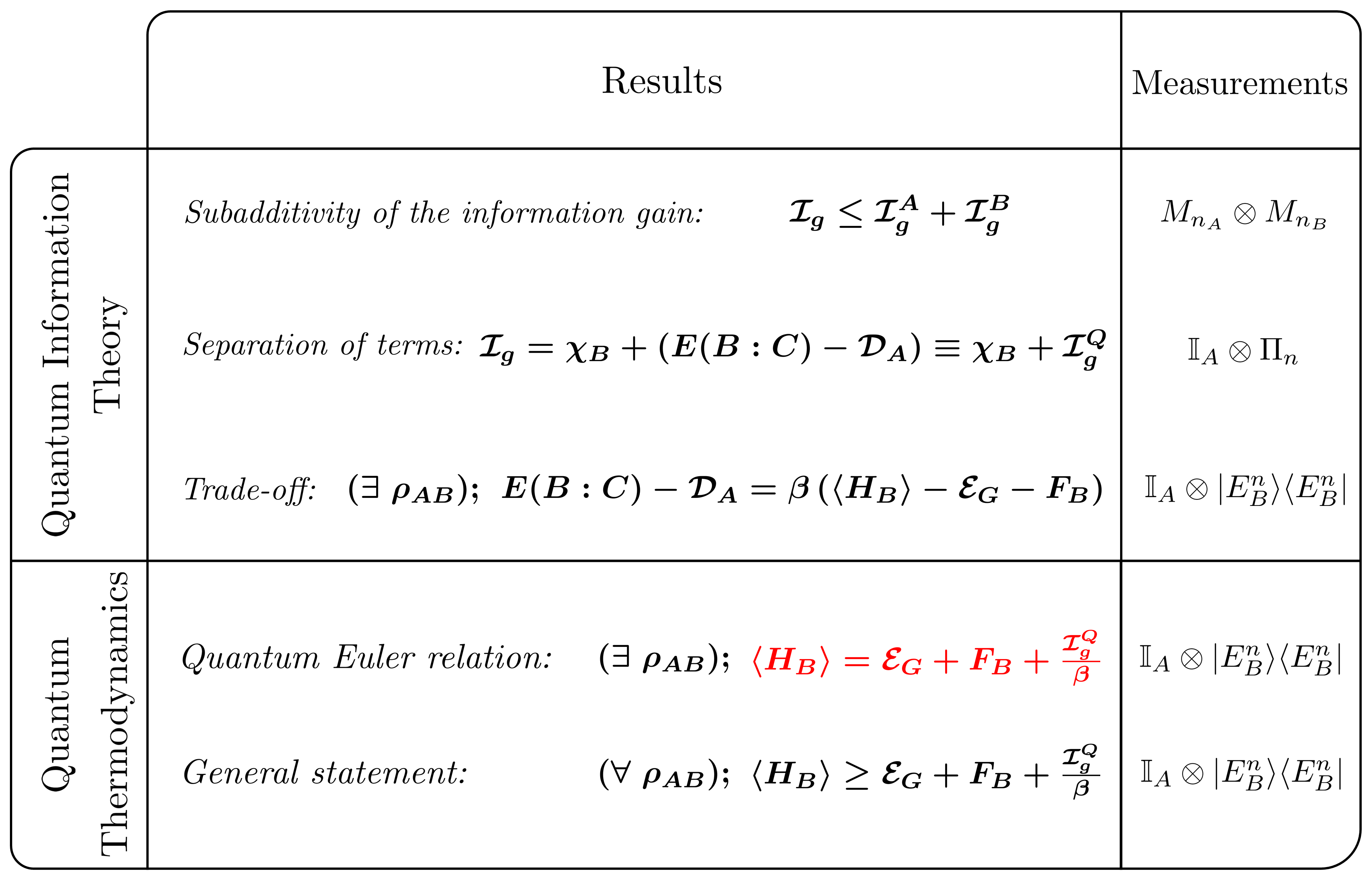}
	\caption[]{\label{tab1}Table summarizing the results of the present manuscript.}
\end{figure}

Remarkably, combining the separation of terms (quantum and classical) and the thermodynamic bounds on $\mc{I}_g$, we arrived at a fundamental identity that is valid for a large class of states. From an information theoretic perspective, we observed that the identity represents a general trade-off relation between genuine quantum correlations within a bipartite state ($\mc{S}$) and the entanglement that one of its partitions ($A$ or $B$) shares with the surrounding environment ($C$). Interestingly, from a thermodynamics perspective, the identity represents a novel quantum formulation of the Euler relation. In this relation, the internal energy is expressed as a function of the ergotropy and the quantum information gain, directly displaying the role of quantum correlations as a thermodynamic resource, and the thermodynamic significance of the quantum information gained through local measurements. A table summarizing the statements derived throughout the paper can be found in Fig.~\ref{tab1}.

Our results highlight the importance of the Groenewold information gain $\mc{I}_g$ as a quantifier for the information output from a quantum measurement and its effects on the state we are probing. Interestingly, the quantum Euler relation can be used in order to estimate the global ergotropy (the ergotropy plus the bound ergotropy) without performing an optimization over all possible unitary operations (cf.~\myeqref{gergo}), and without acting globally on many copies of the state of $\mc{S}$ (to evaluate $\mc{E}_b$). In such cases, we only have to apply local projective energy measurements, on a single partition of $\mc{S}$, in order to compute the global ergotropy for the state of $\mc{S}$. Finally, the main inequalities we presented in this work can be further examined in various scenarios to determine the general class of states for which the quantum Euler relation and the trade-off expression are valid. We leave this analysis for future work on the topic. 

\appendix

\vspace{6pt} 

\funding{S.D. acknowledges support from the U.S. National Science Foundation under Grant No. DMR-2010127. This research was supported by grant number FQXi-RFP-1808 from the Foundational Questions Institute and Fetzer Franklin Fund, a donor advised fund of Silicon Valley Community Foundation (SD).}

\acknowledgments{This work was conducted as part of the Undergraduate Research Program (K.W.) in the Department of Physics at UMBC.}

\conflictsofinterest{The authors declare no conflict of interest.} 

\bibliography{max_info} 

\end{document}